\documentclass[aip,apl,reprint,superscriptaddress]{revtex4-1}

\usepackage{graphicx}
\usepackage{SIunits}

\def\figref#1{Fig.~\ref{#1}}

\begin{document}


\title{Large group delay in a microwave metamaterial analogue of electromagnetically induced transparency} 



\author{Lei Zhang}
\email[Electronic mail: ]{mywaters@iastate.edu}
\affiliation{Ames Laboratory---U.S.~DOE, and Department of Physics and Astronomy, 
             Iowa State University, Ames, Iowa 50011, USA}
             
\author{Philippe Tassin}
\affiliation{Ames Laboratory---U.S.~DOE, and Department of Physics and Astronomy, 
             Iowa State University, Ames, Iowa 50011, USA}
\affiliation{Department of Applied Physics and Photonics, Vrije Universiteit Brussel, 
             Pleinlaan~2, B-1050~Brussel, Belgium}

\author{Thomas Koschny}
\affiliation{Ames Laboratory---U.S.~DOE, and Department of Physics and Astronomy, 
             Iowa State University, Ames, Iowa 50011, USA}
\affiliation{Institute of Electronic Structure and Lasers (IESL), FORTH, and Department of Material Science and Technology,
             University of Crete, 71110 Heraklion, Crete, Greece}

\author{Cihan Kurter}
\affiliation{Department of Physics, Center for Nanophysics and Advanced Materials,
             University of Maryland, College Park, Maryland 20742-4111, USA}
             
\author{Steven M.\ Anlage}
\affiliation{Department of Physics, Center for Nanophysics and Advanced Materials,
             University of Maryland, College Park, Maryland 20742-4111, USA}

\author{C.\ M.\ Soukoulis}
\affiliation{Ames Laboratory---U.S.~DOE, and Department of Physics and Astronomy, 
             Iowa State University, Ames, Iowa 50011, USA}
\affiliation{Institute of Electronic Structure and Lasers (IESL), FORTH, and Department of Material Science and Technology,
             University of Crete, 71110 Heraklion, Crete, Greece}

\date{\today}

\begin{abstract}
We report on our experimental work concerning a planar metamaterial exhibiting classical electromagnetically induced transparency (EIT). Using a structure with two mirrored split-ring resonators as the dark element and a cut wire as the radiative element, we demonstrate that an EIT-like resonance can be achieved without breaking the symmetry of the structure. The mirror symmetry of the metamaterial's structural element results in a selection rule inhibiting magnetic dipole radiation for the dark element, and the increased quality factor leads to low absorption ($<10\%$) and large group index (of the order of 30).
\end{abstract}


\maketitle 

Electromagnetically induced transparency (EIT) is a quantum mechanical process that appears in coherently laser-driven atomic media, such as metal vapours and quantum dot ensembles.\cite{Fleischhauer-2005} Due to the existence of a dark superposition state with vanishing electric dipole moment at resonance, the medium develops a very narrow transmission window in the broader Lorentzian-like absorption peak associated with
the transition to the excited state. Experiments have now established the slowdown of electromagnetic pulses by 7 orders of magnitude.\cite{Kasapi-1995,Hau-1999} Recently, three different groups have independently proposed metamaterials that mimick the response of EIT media;\cite{Zhang-2008,Tassin-2008b,Zheludev-2008} these media achieve their EIT-like response through the interaction of radiation with purely classical, subwavelength electromagnetic resonators with no pump source required. 

Classical EIT media also rely on the use of ``dark'' elements, i.e., resonant elements with vanishing dipole interaction to the external electromagnetic field. Nevertheless, it is possible to distinguish between two different mechanisms to excite the dark resonator. Some researchers have introduced a small coupling to the dark resonator. This unequal coupling to bright and dark elements results in the the typical asymmetrically shaped resonance that is also observed for Fano resonances.\cite{Zheludev-2008,Papasimakis-2009,Chen-2009,Kim-2010} Others have introduced an additional resonant element that carries an electric dipole commensurate with the external field and couples quasi-statically to the dark resonator;~\cite{Zhang-2008,Tassin-2008b,Tassin-2009c,Liu-2009,Singh-2009} these works typically need an asymmetric structure to achieve nonzero coupling strength.

\begin{figure}[b!]
\includegraphics{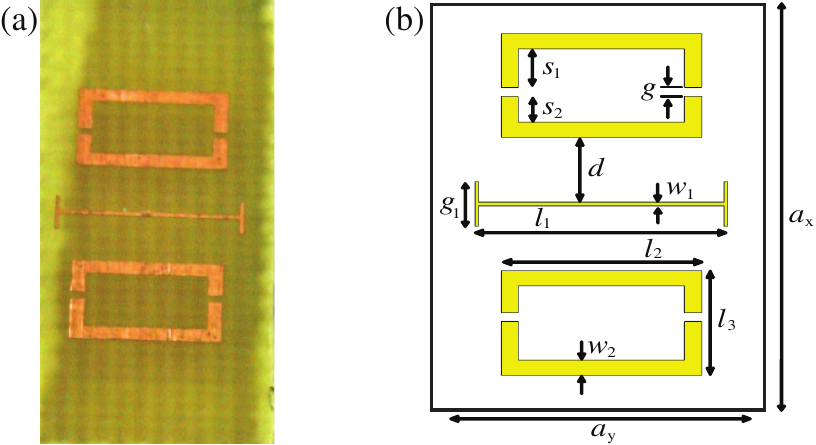}
\caption{(a)~Photograph of the sample.
         (b)~Schematic representation of our structure. The dimensions of the sample are $a_\mathrm{x} = \unit{21}{\milli\meter}$ and $a_\mathrm{y} = \unit{9}{\milli\meter}$. The incident wave is polarized along the $y$ axis. Feature sizes are: cut-wire length $l_1 = \unit{7.3}{\milli\meter}$, cut-wire width $w_1 = \unit{0.1}{\milli\meter}$, wire cap width $g_1 = \unit{1.5}{\milli\meter}$, SRR length $l_2 = \unit{5.8}{\milli\meter}$, width $l_3 = \unit{3.5}{\milli\meter}$, SRR wire width $w_2 = \unit{0.5}{\milli\meter}$, SRR gap $g = \unit{0.3}{\milli\meter}$ (note that the gap is not located at the center of the SRRs due to the nonuniform electric field distribution inside the waveguide), $s_1 = \unit{1.35}{\milli\meter}$, and $s_2 = \unit{1.15}{\milli\meter}$. The distance between the SRRs and the cut wire is $d = \unit{2.2}{\milli\meter}$.}
\label{fig.structure}
\end{figure}

\begin{figure}[t!]
\includegraphics{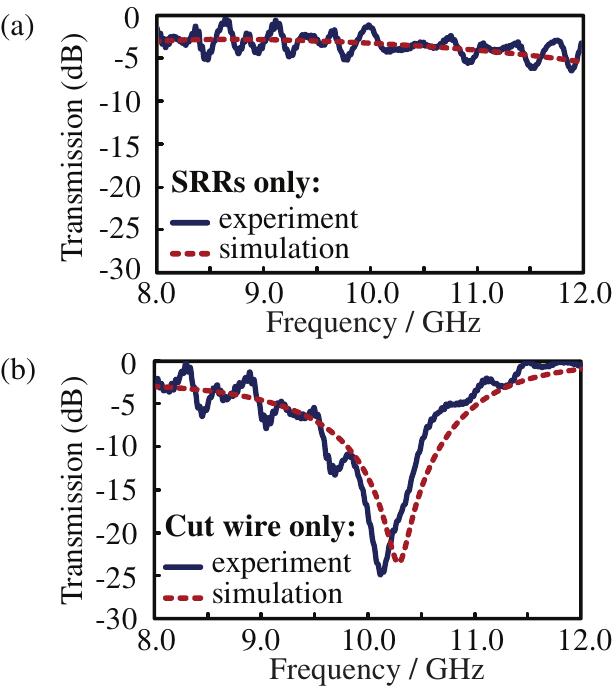}
\caption{Measured and simulated transmittance spectra for (a)~the sample with only the SRRs and (b)~the sample with only the cut wire, when illuminated with the fundamental waveguide mode. The spectra for the cut wire show the electric dipole coupled resonance and the absence of any features in the spectra for the SRRs shows that the magnetic dipole resonances are ``dark.''}
\label{fig.elements}
\end{figure}

In this letter, we follow the second approach to obtain an EIT-like response in a metamaterial, but we deliberately avoid asymmetry in the dark resonator. A photograph of the structure under investigation is shown in \figref{fig.structure}(a) and its geometry is drawn in \figref{fig.structure}(b). For the radiative resonator, we use a finite-length wire (cut wire) oriented parallel to the direction of the electric field of the incident waveguide mode. The cut wire can be directly excited by the external field since it can couple to its electric dipole moment. For the dark element, we use the magnetic dipole mode of two mirrored double-gap split-ring resonators (SRRs). If the incident wave were a plane wave, it would suffice to ensure that each SRR has a mirror plane parallel to the electric field vector in order to avoid coupling of the electric field of the incident waveguide mode to the magnetic resonance.~\cite{Katsarakis-2004} However, since our measurements will be performed inside an X-band waveguide, the electric field intensity is not uniform along the $x$ axis and we have to move the gaps to achieve the same effect. We have designed the position of the gaps in such a way that the induced electric currents along the two arms of each SRR are identical---and thus result in vanishing magnetic moment for each SRR---when directly excited by the fundamental mode of the waveguide. This design consideration guarantees that the magnetic mode of the SRRs cannot be directly excited by the incident electromagnetic wave in the waveguide and, therefore, that the SRRs act as dark elements. In addition, we use two SRRs to restore the symmetry of the structure. The magnetic dipole modes of both split rings will hybridize into two modes: one mode where the circular currents flow in the same direction and one where the currents flow in opposite directions. From symmetry arguments, it is clear that only the latter mode can be coupled to the cut wire. The advantage of this setup is that the total magnetic moment of the dark resonance vanishes. Indeed, because of the symmetry of the electromagnetic fields of the fundamental waveguide mode, the currents in both SRRs are opposite and therefore also their magnetic moments are opposite. There is a residual magnetic quadrupole moment because of the relative displacement of the opposing dipole moments, but a quadrupole has extremely low radiation efficiency. All this means that the dark element has neither an electric nor a magnetic dipole moment, effectively decreasing its radiation resistance in the quasi-static circuit perspective. This property will be supported by the absence of any spectral feature in the absorption spectrum of the dark elements alone in our experimental data presented below.

\begin{figure}[t!]
\includegraphics{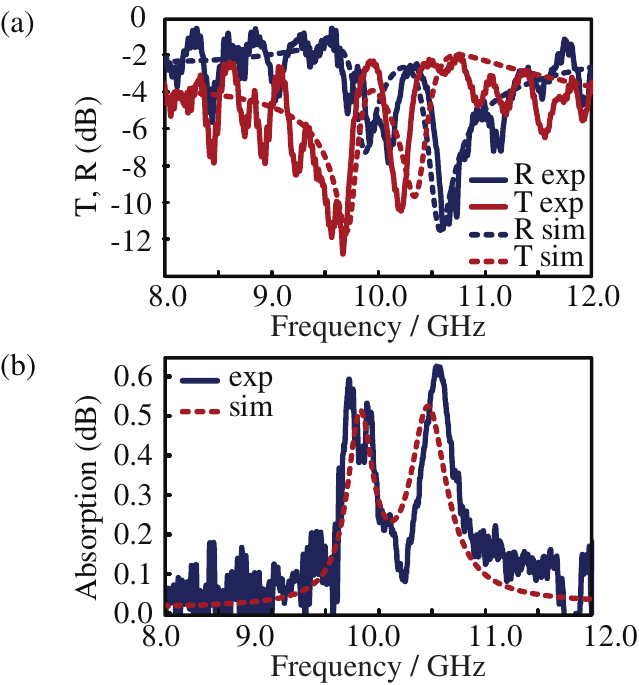}
\caption{Measured and simulated spectra for the sample with both the cut wire and the SRRs, when illuminated with the fundamental waveguide mode. (a)~Transmittance and reflectance. (b)~Absorption. All spectra exhibit the typical features of EIT-like response: a narrow transmission window with low absorption inside a broader resonance.}
\label{fig.spectrum}
\end{figure}

We continue with our experimental and numerical results for this structure with a cut wire coupled to a double SRR structure as shown in \figref{fig.structure}(a). The two SRRs and the cut wire are patterned on a single-sided copper clad FR-4 board of thickness \unit{0.8}{\milli\meter}. All the dimensions of the sample can be found in the caption of \figref{fig.structure}(b). The dielectric constant of the FR-4 board is $\epsilon_\mathrm{r}=4.5+0.15\mathrm{i}$. The samples are measured inside a 12-inch (\unit{30.48}{\centi\meter}) long WR-90 waveguide and the scattering parameters are measured with a Hewlett-Packard E8364 network analyzer. All the numerical results were obtained with the time-domain electromagnetics solver CST Microwave Studio with perfect electric conducting boundaries modelling the waveguide, except if mentioned otherwise. As we have pointed out in earlier work,\cite{Tassin-2008b} three criteria must be fulfilled: (i)~The dark and radiative resonators must share the same resonance frequency; here, we have designed the split rings and the cut wire to be resonant at $\unit{9.9}{\giga\hertz}$. (ii)~The dark resonator may not couple to the incident field. We have checked this requirement numerically and experimentally with an additional FR-4 substrate that contains only the SRRs. The results plotted in \figref{fig.elements}(a) show no structure in the transmittance, confirming the darkness of the double SRR. (c)~The radiative element must have a significantly larger loss than the dark element. The quality factor of the radiative element can be evaluated using a board with only the cut wire present; from the FWHM of the experimental absorption peak and from a numerical eigenmode analysis (obtained with COMSOL Multiphysics), we find that the Q-factor of the cut wire is approximately $Q_\mathrm{wire} = 6$. It is difficult to determine the Q-factor of the dark element experimentally because of the absence of electric and magnetic dipole moment, but we find $Q_\mathrm{dark} = 56$ from the eigenmode analysis. The quality factors of the radiative and dark resonances thus differ by about one order of magnitude, which is sufficient to observe the classical EIT effect.

\begin{figure}[t!]
\includegraphics{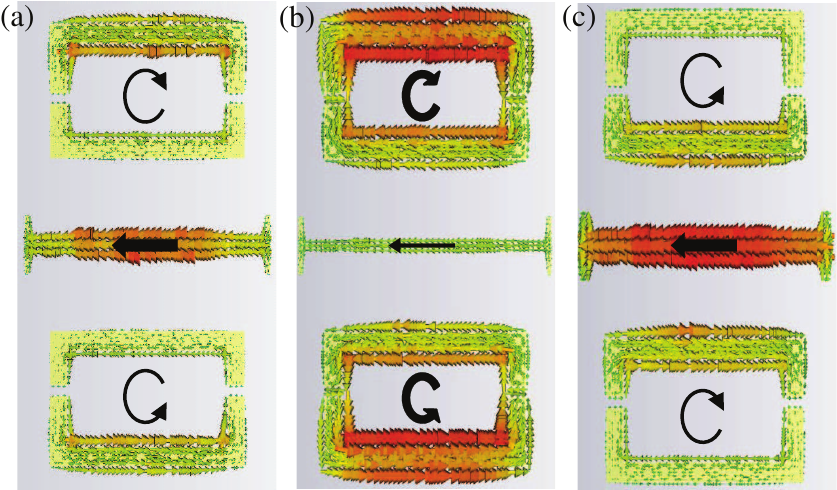}
\caption{Current distributions in the structure at (a)~the leftmost absorption peak ($f = \unit{9.83}{\giga\hertz}$), (b)~at the absorption minimum ($f = \unit{10.05}{\giga\hertz}$), and (c)~at the rightmost absorption peak ($f = \unit{10.48}{\giga\hertz}$).}
\label{fig.currents}
\end{figure}

In \figref{fig.spectrum}(a)-(b), we present the transmittance, reflectance, and absorption spectra for the complete metamaterial structure. We recover the typical features of classical EIT: a narrow transmission window and a narrow region of low absorption ($A < 10\%$) inside a broader absorption envelope that is related to the electric dipole resonance of the wire [compare with \figref{fig.elements}(b)]. Our numerical results are in good agreement with the experimental measurements. Further evidence of EIT-like behavior can be obtained from the simulations. In \figref{fig.currents}, we plot the surface current distributions. At the absorption peaks [Figs.\ref{fig.currents}(a) and (c)], most of the current is distributed along the cut-wire resonator. At at the center frequency of $f = \unit{10.05}{\giga\hertz}$ [Fig.\ref{fig.currents}(b)], there is much less current in the wire due to the destructive interference of the normal modes of the wire-SRR system. However, at the latter frequency there are strong currents in the two SRRs, which does not lead to large absorption because of the high quality of the dark resonance. We again point out the importance of the vanishing dipole moment for the obtained reduction of the absorption.

\begin{figure}[t!]
\includegraphics{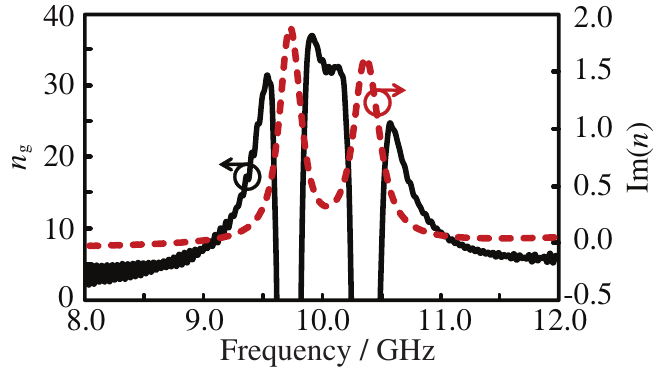}
\caption{Group index and imaginary part of the phase index as obtained from the scattering parameters. At the center frequency, we measure a group index of more than 30, wheras the imaginary part of the index of refraction again confirms the low loss in the sample.}
\label{fig.retrieval}
\end{figure}

The large currents, nevertheless, are related to a strong resonance and one can therefore expect strong dispersion. We use the parameter retrieval procedure developed by Smith~\emph{et~al.}\cite{Smith-2005} for a metamaterial of thickness $a_\mathrm{z} = \unit{2}{\milli\meter}$ to obtain the refractive index $n$. \figref{fig.retrieval} displays the group index defined by $n_\mathrm{g} = n + \omega\,\mathrm{d}n/\mathrm{d}\omega$ and the imaginary part of the phase index. Inside the transparency window, there is strong dispersion and a significantly increased group index of over 30. It is useful to compare the bandwidth-delay product (BDP) of our metamaterial ($\Delta{f}\Delta{t} \approx 0.1)$ to that of quantum-mechanical EIT (e.g., $\Delta{f}\Delta{t} \approx 10$ in Ref.\cite{Hau-1999}); the BDP of our classical system is two orders of magnitude smaller, but we have lower absorption. We can stack several metamaterial layers to increase the BDP, but this approach will eventually be limited by the absorption. It is therefore interesting to pursue other ways to reduce the loss of the dark resonance in addition to the inhibition of magnetic dipole radiation proposed here. 

In conclusion, we have designed and studied a planar metamaterial that exhibits electromagnetically induced transparency around \unit{10}{\giga\hertz}. The dark resonator is implemented by two mirrored split-ring resonators, resulting in vanishing electric and magnetic dipole moments for the dark resonance. The advantage of this approach is the increased quality factor of the dark resonance. Our experimental and numerical results show that this leads to a narrow transparency window with absorption less than $10\%$ together with a group index of over 30.

Work at Ames Laboratory was partially supported by the Department of Energy (Basic Energy Sciences) under Contract No.\ DE-AC02-07CH11358 (computational studies). This work was partially supported by the U.S.\ Office of Naval Research, Award No.\ N000141010925 (synthesis and characterization of samples), and the European Community FET project PHOME, Contract No.\ 213310 (theoretical studies). Work at Maryland was supported by the U.S.\ Office of Naval Research, Grant No.\ N000140811058, and CNAM. P.~T.~acknowledges a Fellowship from the Belgian American Educational Foundation.

%

\end{document}